\documentclass[
 preprint,prx,
 amsmath,amssymb]{revtex4-2}

\usepackage{CJK}
\usepackage{graphicx}
\usepackage{dcolumn}
\usepackage{bm}
\usepackage{float}
\usepackage{color}

\usepackage{enumerate}
\usepackage{xr}
\makeatletter
\newcommand*{\addFileDependency}[1]{
\typeout{(#1)}
\@addtofilelist{#1}
\IfFileExists{#1}{}{\typeout{No file #1.}}
}\makeatother

\newcommand{\stfcpaolo}{Scientific Computing Department, STFC UKRI, Daresbury Laboratory, Warrington WA4 4AD, United Kingdom}
\newcommand{\stfcgilberto}{Scientific Computing Department, STFC UKRI, Rutherford Appleton Laboratory, Didcot  OX11 0QX, United Kingdom}
\newcommand{\ect}{European Centre for Theoretical Studies in Nuclear Physics and Related Areas
(ECT*), Bruno Kessler Foundation, Trento, Italy}
\newcommand{\tifpa}{Trento Institute for Fundamental Physics and Applications (TIFPA), Istituto Nazionale di Fisica Nucleare, Italy}
\newcommand{\epsic}{Electron Physical Science Imaging Center, Diamond Light Source Ltd., Didcot OX11 0DE, United Kingdom}
\newcommand{\oxf}{Department of Materials, University of Oxford, Oxford OX1 3PH, United Kingdom}

\begin{document}

\title{The role of primary and secondary electrons in scanning transmission electron microscopy of hybrid perovskites: the CsPbBr$_{3}$ case.}

\author{P. E. Trevisanutto}
\email{paolo.trevisanutto@stfc.ac.uk}
\affiliation{\stfcpaolo}

\author{S. Taioli}
\affiliation{\ect}
\affiliation{\tifpa}
\author{M. Dapor}
\affiliation{\ect}
\affiliation{\tifpa}
\author{C. S. Allen}
\affiliation{\epsic}
\affiliation{\oxf}
\author{G. Teobaldi}
\affiliation{\stfcgilberto}

\begin{abstract} 
High-resolution imaging has revolutionised materials science by providing detailed insights into the atomic structures of materials. Electron microscopy and spectroscopy rely on analysing backscattered and transmitted electrons as well as stimulated radiation emission to form structural and
chemical maps. These signals contain information about the elastic and inelastic electron scattering
processes within the sample, including collective and single electron excitations such as plasmons,
inter- and intra-band transitions. In this study, ab initio and Monte Carlo simulations were performed 
to investigate the behaviour of high-energy primary and secondary electrons in scanning
transmission experiments on CsPbBr$_3$ nanosamples. CsPbBr$_3$ is a perovskite material known for
its high photoluminescence quantum yield, making it promising for applications in light-emitting devices and solar cells. This study investigates and estimates the reflection and transmission of primary and secondary electrons based on their kinetic energy, sample thickness and electron affinity.
The spatial distribution and energy spectra of the secondary electrons are also analysed and calculated to understand their generation depth and energy dynamics. These findings provide a theoretical framework for the study of charge transport in perovskites and can help to optimise scanning microscopy techniques for the imaging and characterisation of advanced materials.
\end{abstract}


\maketitle 
\section{Introduction}

Among the significant advances in analytical techniques for materials science, atomic resolution imaging is one of the most impactful. These methods have enabled detailed visualisation of the structures and morphology of nanomaterials, crystals and metal surfaces, as well as biological samples and tissues \cite{Review_STEM_prog_surf,Nellist2019}.

In transmission electron microscopy (TEM), a beam of high-energy electrons is passed through a very thin sample ($\sim$ 100~nm) and interference between elastically scattered electrons is exploited to form an image. This approach enables high-resolution imaging down to the atomic resolution ($\sim$ 0.1 nm) and allows the precise visualisation of atomic arrangements. TEM microscopes typically operate at electron energies of 80-300 keV, although atomic resolution has already been demonstrated at 30 keV \cite{PhysRevLett.114.166102,10.1063/5.0143684} and even 15 keV \cite{SASAKI201450}.

In a scanning electron microscope (SEM), a focused beam of electrons is raster scanned across a sample, and nm resolution images of the surface are formed from the analysis of emitted secondary electrons (SE) and backscattered electrons (BSE). The resolution in a SEM is largely limited by the interaction volume of the detected radiation, which, in most cases, precludes the need for incident electron probes smaller than $\sim$ 1 nm. SEM microscopes typically operate at electron beam kinetic energies of 1-30 keV.
Images in a SEM are generally formed by collecting secondary or backscattered electrons that originate from the top surface of the sample. Performing in a low voltage mode, SEM in transmission mode has advantages over TEM imaging (albeit with lower spatial resolution) due to the relatively low cost and simplicity of the SEM instrument \cite{KLEIN2012297}and has applications in both the physical \cite{10.1093/micmic/ozac010} and biological \cite{kuwajima2013automated} sciences.

Scanning transmission electron microscopy (STEM) combines elements of SEM and TEM. A finely focused electron beam is raster scanned across a thin sample ($\sim$ 100 nm), and scattered electrons are detected to form an image. At high incident electron energy and for thin samples, the interaction volume is small and the microscope resolution is primarily limited by the size of the incident electron probe, which, with modern aberration-corrected optics, can reach sub-Ångström dimensions. Together with other emissions such as Auger electrons which are used to characterise the surface elemental compositions and X-rays applied to determine bulk crystal structures, the SEM, TEM and STEM represent the main techniques to analyse chemical composition and structure of materials \cite{TAIOLI2010237}.

In high-resolution STEM and TEM experiments, a projection image is acquired, and it is challenging to obtain information on a sample variation along the beam direction. By placing SE detectors both before and after the sample within a STEM instrument, the surface sensitivity of SE detection can be exploited to determine the structure on both entrance and exit surfaces of a thin sample \cite{Atomic_Resolution_Secondary,Se_Damage_cambridge_2022}. This approach, to some extent, overcomes the projection limitation of the STEM. 

A recent review \cite{microscopy_review_se} has highlighted the applications for studying the dynamics and transient phenomena of surface reconstructions, exsolution of catalysts, lunar and planetary materials and the mechanical properties of 2D thin films utilising the simultaneous SE-STEM imaging.

Furthermore, inelastic scattering events within the sample, including SE generation, \cite{sec_induced_mech} are a primary cause of electron beam-induced sample damage and need to be well understood to develop new approaches for imaging beam-sensitive materials.
Motivated by the potential for transmitted SE imaging described in the above works, we construct a theoretical framework to calculate SE generation and emission in transmission through an industrially relevant metal-halide perovskite sample.  

Any electron microscopy technique relies on electron scattering in the sample, which includes both elastic and inelastic processes. Both spectroscopic and SE imaging are products of inelastic scattering processes within the sample. These mechanisms are driven by energy loss phenomena, including single-particle electron excitations, collective atomic modes such as plasmons, electron-phonon and electron-polaron interactions, and Auger decay. 
Part of the energy loss contributes to the generation of a secondary electron cascade and, upon emission from the surface, to the secondary electron yield (SEY). Secondary electrons appear in the electron energy spectrum as a structured broad peak with energies below 50 eV. The emission of secondary electrons in SEM plays a decisive role in the formation of image contrast \cite{dapor_ciappa} and can also lead to unwanted charge of the sample \cite{Taioli_2023_lisadagliocchiblu}.

Electron energy loss phenomena are influenced by two factors, namely the properties of the beam (i.e. kinetic energy, radiation type, and charge state) and the material properties, in particular the electronic excitation spectrum. The latter can be described by the so-called energy loss function (ELF), which can be obtained either from optical experiments or from EELS using a single scattering spectrum with the elastic peak and multiple scattering removed \cite{1Egerton} or via a Reverse Monte Carlo (RMC) method \cite{Li2023ImprovedRM}. The ELF is a unique property of the material and is independent of the beam. However, it is difficult to interpret its spectral profile from the experiment alone, as the uniqueness of the ELF is not guaranteed by the fulfilment of the sum rules, i.e. the EELS signal can lead to different ELFs that all fulfil the sum rules \cite{10.3389/fmats.2023.1249517} and thus to different beam stopping properties.
This emphasises the need for more precise theoretical descriptions and predictions of electron energy loss. In this context, ab initio calculations can also be used to determine the dielectric properties \cite{taioli2009electronic}, especially in the low energy loss region.

In this article, the scattering behaviour of high-energy primary electrons (up to 50 keV, which covers the energy range of SEM and low energy TEM operation) and secondary electrons (up to 50 eV) in STEM experiments on CsPbBr$_3$ thin films is theoretically investigated, using a mixed first-principles and Monte Carlo (MC) approach. In particular, linear response time-dependent density functional theory (LR-TDDFT) is used to calculate the ELF, while the MC routine models the charge transport in the solid. We have also investigated the reflection and transmission of primary and secondary electrons as a function of the kinetic energy of incidence and the sample thickness. Because electron microscopy (EM) samples are susceptible to beam-induced charging, we also determined the effects of the Electron Affinity (EA) on electron emission (we recall that the EA is measured from the bottom of the conduction band inside semiconductors and insulators). Finally, we determined the SE spectra and spatial distribution as functions of the initial kinetic energy of the primary beam by calculating the depth of secondary electron generation.

Caesium lead bromide, CsPbBr$_3$, is an inorganic perovskite material that has attracted much attention in optoelectronics due to its high photoluminescence quantum yield (PLQY) with a sharp, narrow-band blue-green light emission \cite{blue1,blu2,blue3,blue4,blue5,blue6}. These properties make it a good candidate for industrial applications, e.g. in light-emitting devices (LEDs).
As CsPbBr$_3$ is a purely inorganic material, it is more stable under heat and less susceptible to degradation under harsh environmental conditions \cite{dur1,dur2,dur3}. 
Although CsPbBr$_3$ has a relatively large bandgap (about 2.3 eV \cite{Ezzeldien2021}), it can be used in tandem solar cells alongside silicon or other materials to cover different parts of the solar spectrum and increase overall efficiency.
In addition, CsPbBr$_3$ exhibits the ``hot carrier effect'' \cite{hotcarrier1,hotcarrier2}, in which excited electrons maintain high energy levels longer than in other materials, which has the potential to further increase the efficiency of solar cells. It has also shown the possibility of trion generation under high-energy laser excitation \cite{Trion_Cao_2020}.

\section{Theory and methods}


\subsection{Energy Loss Function}

The energy loss processes of electrons are described using Ritchie's dielectric theory \cite{1Ritchie57,TAIOLI2024100646}. This approach is based on the knowledge of the macroscopic dielectric function $\overline{\epsilon}(\textbf{q},\omega)$, which can be calculated from first principles using LR-TDDFT or obtained from optical experiments. 
\begin{figure}[hbt!]     
\includegraphics[width=0.5\textwidth]{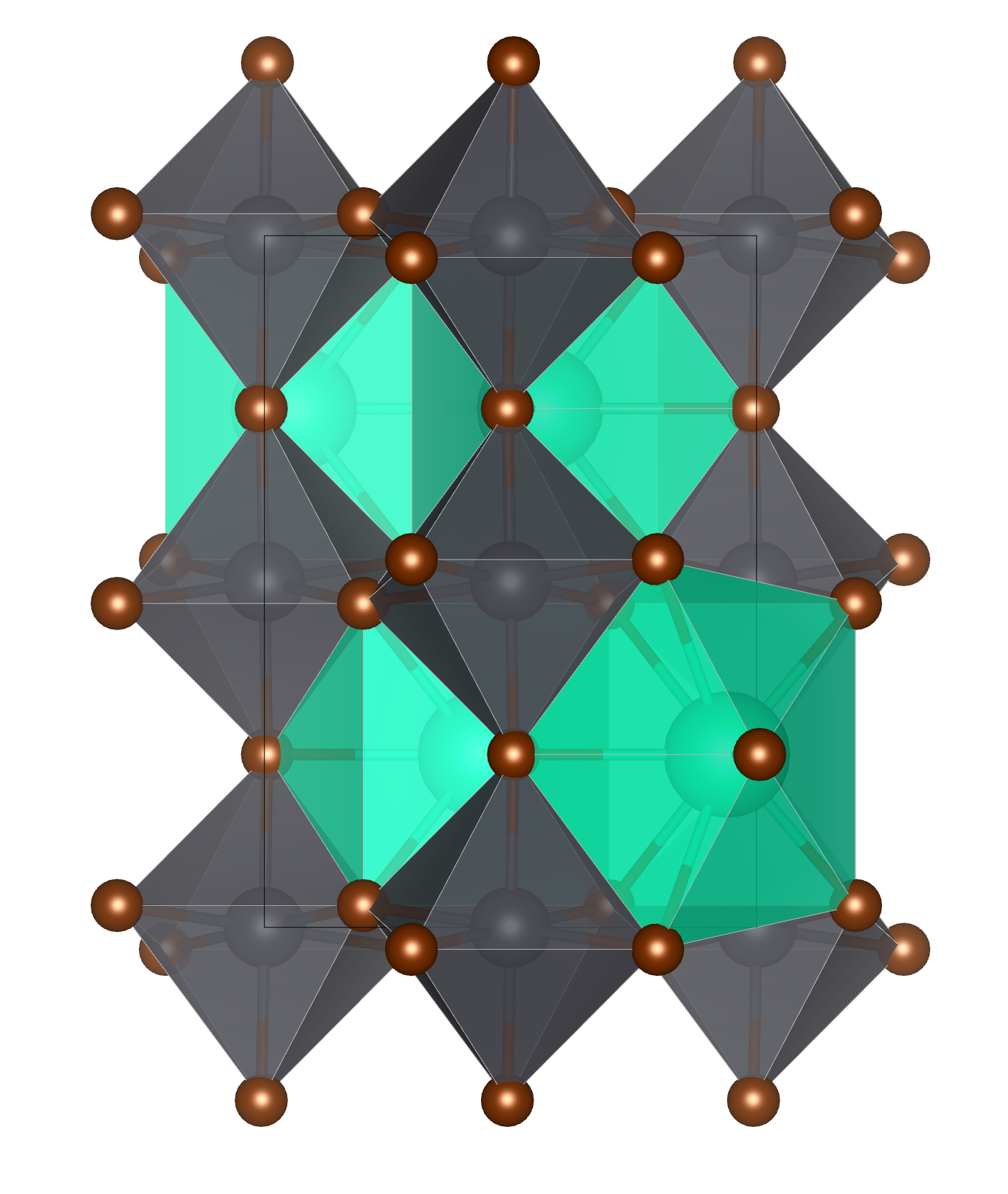}
\caption{Geometric structure of CsPbBr$_3$ (crystallographic space group $Pnma$) optimised by using density functional theory. Br is shown in brown and surrounds Pb (grey) and Cs (green) atoms.}
\label{fig:struct}
\end{figure}
The former allows for the inclusion of momentum dispersion, which is necessary for the calculation of the stopping properties of materials, with the same computational effort as in zero-momentum transfer, whereas the latter approach must be extended to finite momentum by analytical dispersion laws, which are known to be valid only within certain limits \cite{TAIOLI2024100646}.
We have considered the orthorhombic geometric structure of CsPbBr$_3$ \cite{calc_dft_matproj,Ezzeldien2021}, which belongs to the crystallographic space group $Pnma$ and is shown in Fig. \ref{fig:struct}.

The electronic structure in the ground state was relaxed in the framework of density functional theory (DFT) using the generalised gradient approximation (GGA) of the exchange-correlation functional as implemented in the Quantum Espresso (QE) code suite \cite{Giannozzi_2017}.
We have checked the convergence of the DFT parameters given in the literature \cite{esempio,Tomanov2019OnTS,calc_dft_matproj}. We note that the DFT and LR-TDDFT QE calculations performed under the random phase approximation (RPA) neglecting local field effects (RPA-NLFE) were compared with the same calculations using Elk \cite{elk}, an all-electron full-potential linearised augmented plane-wave (LAPW) code and Questaal \cite{PASHOV2020107065}. The RPA-LFE LR-TDDFT calculations were performed with the Liouville-Lanczos turboEELS code \cite{TIMROV2015460}. We have converged our $k$-mesh grid to 21$\times$21$\times$21 points. 

In the LR-TDDFT, the susceptibility function $\chi(\textbf{q},\omega)$ can be obtained by solving the following Dyson-like equation:
\begin{equation}
\chi(\textbf{q},\omega)^{-1}=\chi(\textbf{q},\omega)^{-1}_0 -v_C(\textbf{q})-f_{xc}(\textbf{q},\omega),
\label{susceptibility}
\end{equation}
where $\textbf{q}$ is the momentum transfer vector, $\omega$ is the absorbed energy, $\chi(\textbf{q},\omega)_0$ is the non-interacting (or independent) particle susceptibility calculated from the Kohn-Sham wave functions and  $v_C(\textbf{q})$ is the bare Coulomb potential. In this work, we used the Adiabatic Generalised Gradient Approximation (AGGA) \cite{doi:10.1146/annurev.physchem.55.091602.094449} to calculate the TDDFT kernel $f_{xc}(\textbf{q},\omega)$.
The microscopic dielectric function $\epsilon(\textbf{q},\omega)$ of the material is related to the susceptibility as follows:
\begin{equation}
\epsilon(\textbf{q},\omega)=1-v_C(\textbf{q})\chi(\textbf{q},\omega).
\label{epsilon}
\end{equation}
In a periodic crystal, the microscopic dielectric function is denoted in reciprocal space as $\epsilon_{\textbf{G},\textbf{G'}}(\textbf{q},\omega)\equiv\epsilon(\textbf{q}+\textbf{G},\textbf{q}+\textbf{G'},\omega)$, where $\textbf{G}$ and $\textbf{G'}$ are reciprocal lattice vectors. The macroscopic dielectric function (or dielectric matrix) $\overline{\epsilon}$ is related to the microscopic dielectric function by the following expression:
\begin{equation}
\overline{\epsilon}(\textbf{q},\omega)=\left[\epsilon^{-1}\right]^{-1}_{\textbf{G}=0,\textbf{G'}=0}(\textbf{q},\omega).
\label{LFE}
\end{equation}
If this relationship is applied to the momentum $\textbf{q}\rightarrow0$ in the first Brillouin zone is referred to as the ``optical limit''. The inversion of $\epsilon^{-1}$ with $\textbf{G}$ and $\textbf{G'}$ not equal to zero leads to the inclusion of the so-called local field effects (LFE), which take into account the inhomogeneity of the system, in contrast to the non-local field effect (NLFE), in which only the head of the matrix is inverted.
Using Eq. (\ref{LFE}), the ELF can be written as:
\begin{equation}
    {\mathrm{ELF}}\equiv -\mathrm{Im}[\overline{\epsilon}^{-1}]=\frac{\mathrm{Im}[\overline{\epsilon}]}{\mathrm{Im}[\overline{\epsilon}]^2+\mathrm{Re}[\overline{\epsilon}]^2}.
\end{equation}

To determine the effects of the AGGA correlation, the RPA ($f_{xc}(\textbf{q},\omega)=0$) is also used within LFE and NLFE.
\begin{figure}[hbt!]
\includegraphics[width=0.85\textwidth]{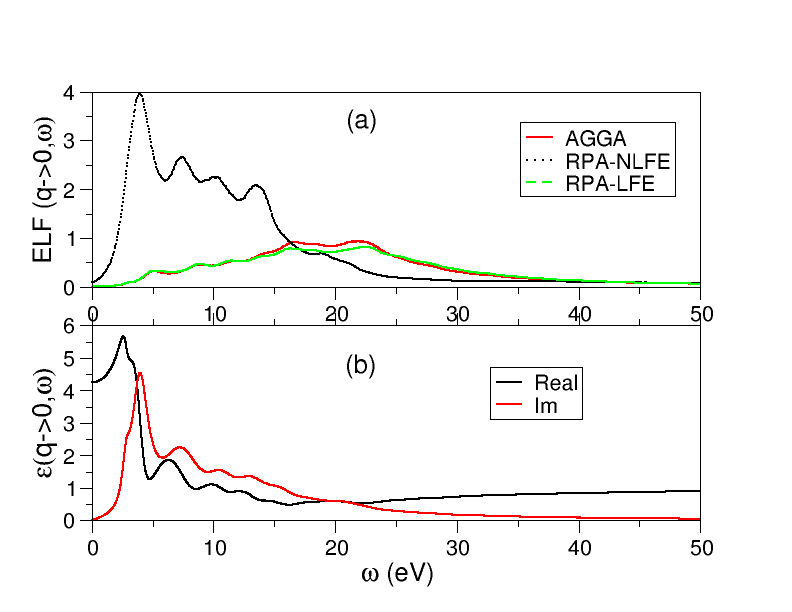}
\caption{(a) ELF$(\textbf{q}\rightarrow 0,\omega)$ of CsPbBr$_3$ in RPA-NLFE (dotted black line), RPA-LFE (dashed green line) and AGGA (red solid line). (b) The real (black solid line) and imaginary (red solid line) parts of the dielectric function are calculated using LR-TDDFT within the AGGA framework.}
\label{fig::elf_tddft}   
\end{figure}
In the upper panel of Fig. \ref{fig::elf_tddft} the ELFs are compared in the context of AGGA, RPA-LFE and RPA-NLFE. As we are currently unable to compare our computer simulations with the experimental results, the peaks are convolved with a typical Gaussian broadening of 0.1 eV.
When RPA-LFE is compared with RPA-NLFE, a dramatic change in the shape of the ELF can be observed due to inhomogeneities. At the same time, the electron correlation in the AGGA kernel slightly increases the intensity in the 15-25 eV range, especially the 16.75 and $\sim$ 22 eV peaks, compared to RPA-LFE. The first less bright peak at 5 eV in the AGGA and RPA-LFE spectra is associated with the transition from the 4$p$ states of the Br atom to the 6$p$ states of Pb. The rich peak structures beyond 6 eV are primarily caused by transitions from the 4$p$ states of the Br atom to a combination of 6$p$ states of Cs and 6$p$ states of Pb.\\
\indent The real and imaginary parts of the dielectric function calculated using LR-TDDFT within the AGGA framework are shown in the bottom panel of Fig. \ref{fig::elf_tddft}. By analysing their spectral behaviour, it can be concluded that the 22 eV peak is associated with a mixed-state plasmon (the real part of the dielectric function remains positive, even if it is almost zero). 
\begin{figure}[hbt!]        
\includegraphics[width=0.8\textwidth]{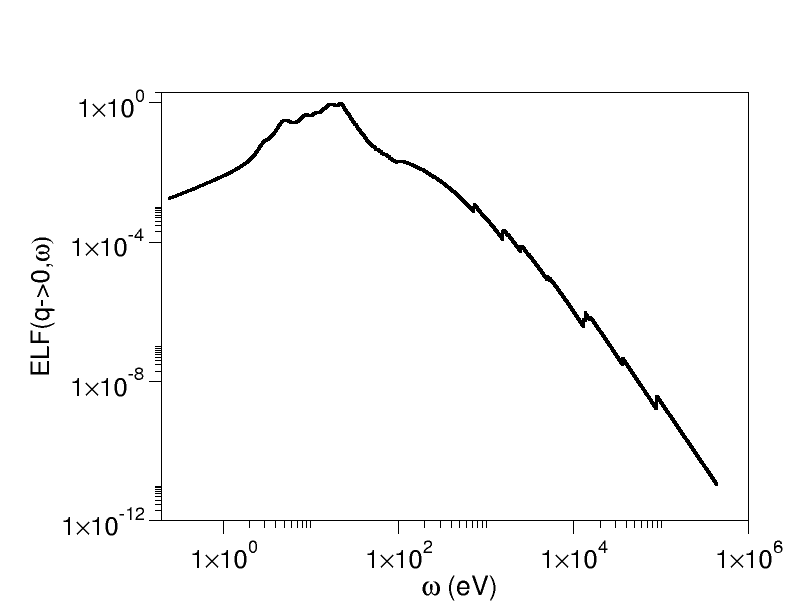}
\caption{ELF of CsPbBr$_3$ in LR-TDDFT LFE AGGA merged with the ELF derived from the experimental form factors of the NIST datasets \cite{NIST}}
\label{fig:elf_nist}
\end{figure}
In view of these results, we decided to use the TDDFT AGGA ELF, which contains the LFE, for our analysis. \\
\indent In Fig. \ref{fig:elf_nist} the TDDFT AGGA LFE ELF is merged with the experimental form factors from the NIST datasets \cite{NIST}. This merging is necessary because the estimation of the high-energy TD-DFT ELF is computationally too demanding because of the numerous excited states that must be included in the calculation. On the other hand, the extension from 100 eV to 400 keV, which is related to the atomic core-level excitations, is needed to accurately determine the total inelastic cross section. 

\subsection{Total inelastic scattering cross section and mean free path}

To calculate the differential inverse inelastic mean free path (DIIMFP), which is used in our MC algorithm, we need to extend the ELF in the optical limit to finite momentum transfer \textbf{q} to account for energy-momentum dispersion. We have opted for the Drude-Lorentz approximation, where the dispersion is typically assumed to be a square law \cite{TAIOLI2024100646} (see Appendix for details).
\begin{figure}[hbt!]
\centerline{\includegraphics[width=0.8\linewidth]{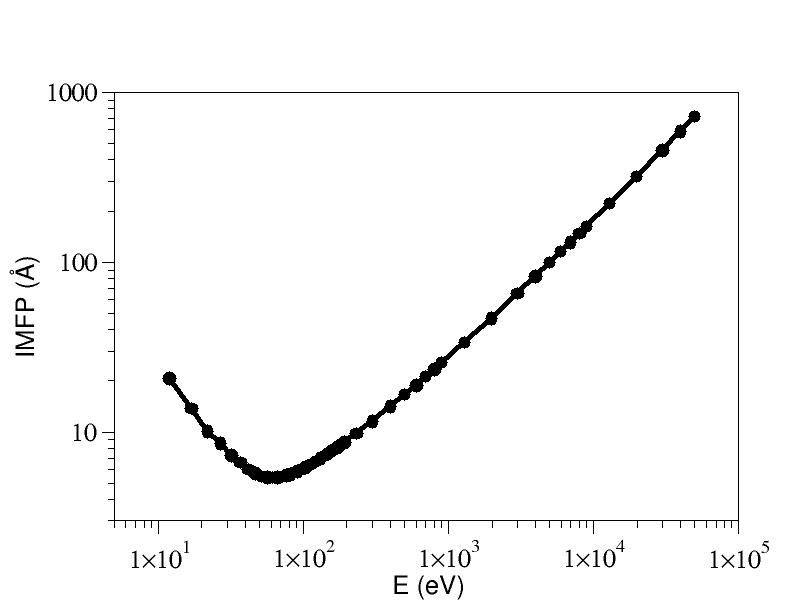}}
\caption{IMFP (\AA) as a function of the kinetic energy of the incident electron (eV), obtained by extending the ELF from TDDFT-AGGA in the energy axis via the NIST datasets and in the momentum transfer axis $\textbf{q}$ according to the quadratic dispersion law \cite{TAIOLI2024100646}. The minimum of the IMFP is 5.34 \AA~for an energy of 57 eV}.
\label{fig:imfp}
\end{figure}
The equation for the inverse inelastic mean free path (IIMFP), from which the total inelastic scattering cross section can be derived, is given by
\begin{equation}
\lambda^{-1}_{\mathrm{inel}}(E)=\frac{1}{E\pi a_0}\int_{W_{\mathrm{min}}}^{W_{\mathrm{max}}}d\omega\int^{q_+}_{q_-}\frac{dq}{q}\mathrm{Im}\left[-\frac{1}{\overline{\epsilon}(q,\omega)}\right]
 \label{eq:imfp_eq},
\end{equation}
where the lower and upper limits of the integral $q_{\pm}= \sqrt{2mE}\pm \sqrt{2m(E-\hbar\omega)}$ are obtained by the energy-momentum conservation laws, $a_0$ is the Bohr radius, $m$ is the electron mass, and according to Ganachaud and Mokrani $W_{\mathrm{min}}$ is the band gap and  $W_{\mathrm{max}}$ corresponds to the maximum energy of the incident electrons \cite{GANACHAUD1995329}. We have tested that the relativistic effects on the IMFP of CsPbBr$_3$ are minimal by performing the calculations with the relativistic extension of Eq. \ref{eq:imfp_eq} \cite{relativistic_shino} (the maximum energy considered in this work is 1/10 of the rest mass of the electron).
Fig. \ref{fig:imfp} shows the IMFP ($\lambda_{\mathrm{inel}}$) for CsPbBr$_{3}$ as a function of the electron kinetic energy.

\subsection{Total elastic scattering cross section and mean free path}

Elastic scattering is accounted for in our MC approach using Mott theory, which we have generalised to account for the presence of multiple scattering between neighbours in the periodic unit cell \cite{3ELSEPA}. In this approach, the differential elastic scattering cross section (DESCS) for an electron scattered by a central atomic potential at an angle $\theta$ is as follows:
\begin{equation}
    \frac{d\sigma_{\mathrm{el}}(E,\theta)}{d\Omega}=\sum_{m,n} e^{i \textbf{q}\cdot \textbf{r}_{m,n}}\left[f_m(\theta)f^*_n(\theta)+g_m(\theta)g^*_n(\theta)\right],
\end{equation}

where $f_m(\theta)$ and $g_m(\theta)$ are the Mott amplitudes of direct and spin-flip scattering for the $m,n$ atoms, which were determined using the relativistic partial-wave expansion method \cite{TAIOLI2024100646} and the modulus of $\textbf{r}_{m,n}=\textbf{r}_m-\textbf{r}_n$ indicates the distance between the $m$th and $n$th atom. $\theta$ and $\Omega$ are the scattering and solid angles, respectively, measured in the laboratory frame.
The total elastic scattering cross section integrated over $\Omega$ results as:

\begin{equation}
    \sigma_{\mathrm{el}}(E)=\int_{\Omega}\frac{d\sigma_{\mathrm{el}}(E,\theta')}{d\Omega'}d\Omega',
\end{equation}
where $d\Omega'=2\pi\sin\theta'd\theta'$.

\section{Monte Carlo results}\label{MC_reels}

Since we have access to the cross sections of elastic and inelastic scattering, our MC algorithm works as follows \cite{Dapor2023,TAIOLI2024100646}: we extract a random number $\mu$ that is sampled from a uniform distribution in the interval $\left[0,1\right]$ and use it to determine the step length that each electron travels within the solid target $\Delta s =-\lambda\ln(\mu)$, where $\lambda$ is the mean free path that includes all elastic ($\lambda^{-1}_{\mathrm{el}}$)  and inelastic ($\lambda^{-1}_{\mathrm{inel}}$) scattering processes and can be calculated from $\lambda^{-1}=\lambda^{-1}_{\mathrm{el}}+\lambda^{-1}_{\mathrm{inel}}$.\\
\indent Finally, the MC method selects an elastic or inelastic scattering event by comparing another random number $\mu'$, which is sampled from a uniform distribution in the range [0, 1], with the corresponding probability. If $\mu' \le \lambda_{\mathrm{el}}^{-1}/\lambda^{-1}$, it is an elastic scattering event that leads to an angular deviation of the electron's trajectory; otherwise, the electron suffers an inelastic interaction that leads to both energy loss and (typically small) angular deviation \cite{Dapor2023}. We note that to escape from the surface of the sample, the electrons must overcome the barrier at the interface between the vacuum and the solid, i.e. they must fulfill the condition
$E\cos^2(\vartheta)\geq \mathrm{EA}$
where $\vartheta$ is the angle formed between the surface normal and the intersection of the electron trajectory on the surface, and $E$ is the kinetic energy of the electrons. In addition, the transmission from solid to vacuum is modeled by a transmission coefficient that depends on the EA of the material \cite{TAIOLI2024100646}.

\subsection{Electron energy loss spectrum}\label{Reels_and_secondary}

\begin{figure}[hbt!]
\centering
\includegraphics[width=0.8\linewidth]{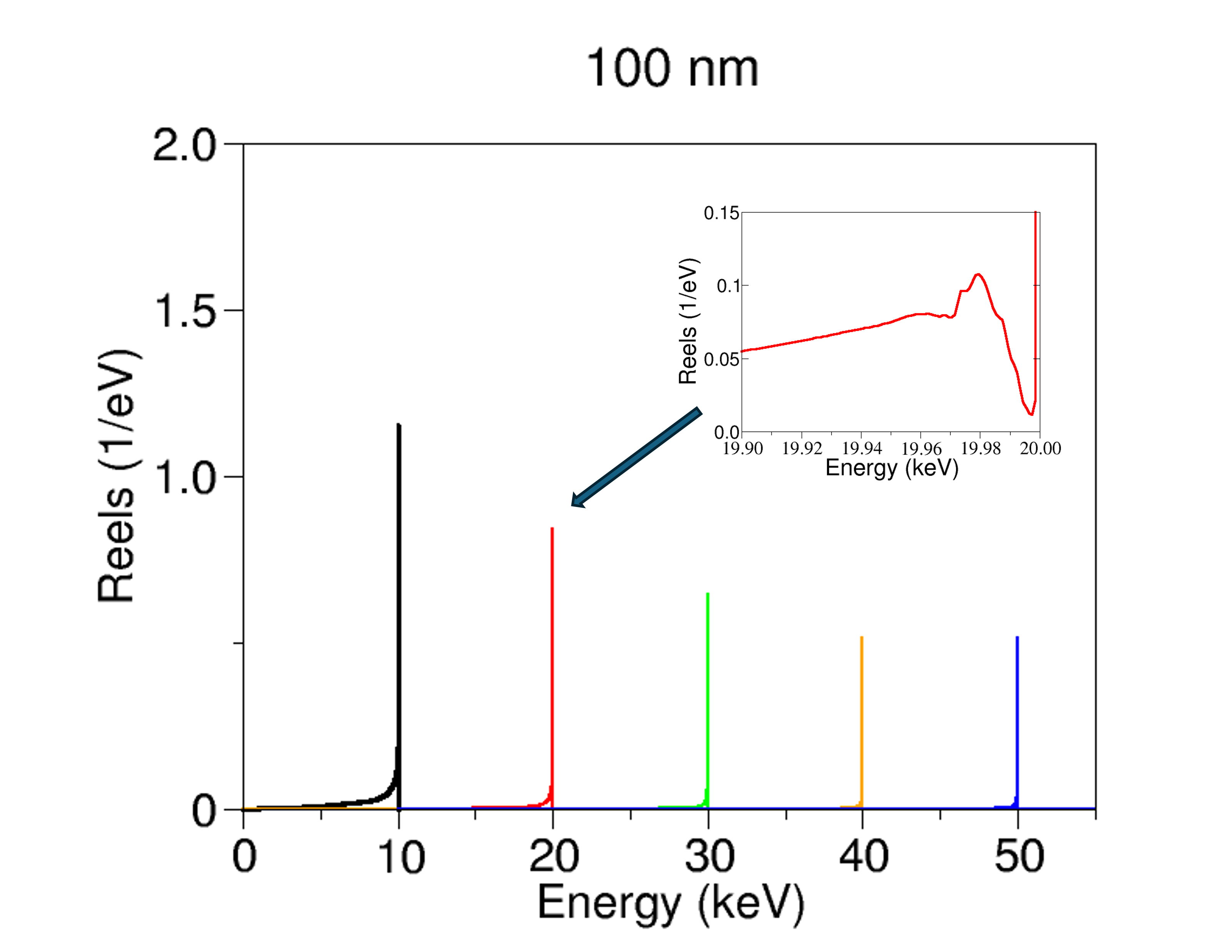}
\caption{\label{EELS} Monte Carlo simulation of the energy loss spectrum of electrons emerging from a 100 nm thick CsPbBr$_{3}$ sample for different kinetic energies of the primary beam: $10$ (black line), $20$ (red line), $30$ (green line), $40$ (orange line) and $50$ (blue line) keV. The inset shows the plasmon peak for a 20 keV electron beam.}
\end{figure}

Using the MC approach described above, we have calculated the reflection electron energy loss spectrum (REELS) \cite{Ding_primary} 
for different initial energies of the incident electrons (black solid line for 10 keV kinetic energy, red line for 20 keV, green line for 30 keV, orange line for 40 keV, and blue line for 50 keV), as shown in Fig. \ref{EELS}. 
Part of the electrons of the primary beam can be backscattered with the same energy as the incident energy and represents the so-called elastic (or lossless) peak, which typically also hides substructures due to electron-phonon scattering. We performed all MC calculations with a total number of electrons equal to $10^{10}$ to obtain a good signal-to-noise ratio.\\
\indent The elastic peaks shown in Fig. \ref{EELS} are extremely bright and lie at the initial kinetic energy of the primary beam. The energy loss spectra show several peaks reflecting the ELF lineshape. In Fig. \ref{EELS}, the inset image highlights the first 100 eV of energy loss for an initial impact energy of 20 keV. The shoulders in the 20 keV REELS at $5$, $8$ and $12$ eV from the elastic peak are directly related to single-particle excitations (Br 4$p$ states $\rightarrow$ Pb 6$p$ states and Br 4$p$ states $\rightarrow$ Cs 6$p$ states), which also occur in the ELF. The highest peak is at $22$ eV away from the elastic peak, which is a mixed single-electron/plasmon transition.

\subsection{Backscattered electrons}

To investigate the effects of the film thickness on the reflected electrons, we also calculated the backscattering coefficient $\eta_B$ (i.e. the fraction of backscattered electrons to total electrons) as a function of incident electron energies of $5$, $10$, 15, $20$, $30$, 40 and $50$ keV. 
\begin{figure}[hbt!]
\includegraphics[width=0.8\linewidth]{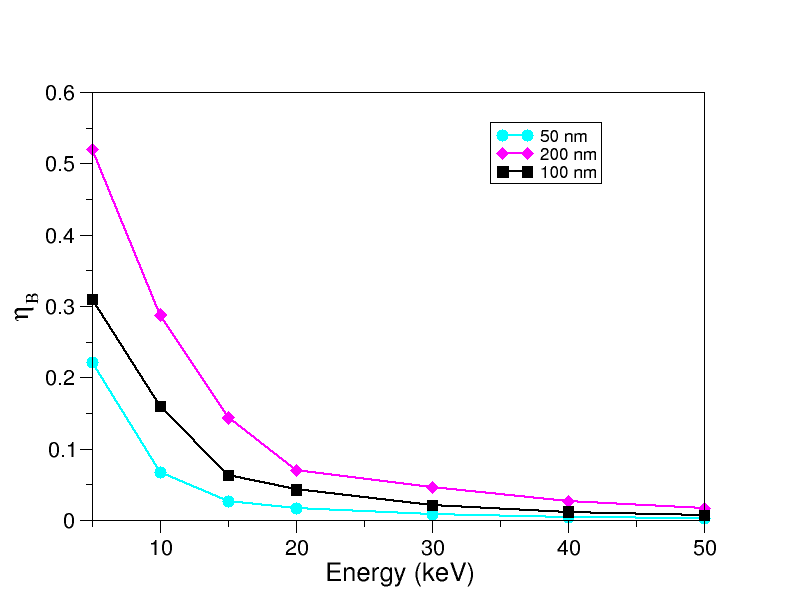}
\caption{Backscattering coefficient ($\eta_B$) for the electrons of the primary beam as a function of their initial kinetic energy (keV).
The results are shown for three different sample thicknesses: 50 nm (cyan-coloured solid line), 100 nm (black solid line) and 200 nm (magenta-coloured solid line). 
}
\label{eta-primary_number}
\end{figure}
In Fig. \ref{eta-primary_number} we show the results for three CsPbBr$_3$ thin films of thickness $50$ (cyan coloured line with full circles), $100$ (black coloured line with full squares) and $200$ nm (magenta coloured line with full diamonds).
The backscattering coefficients for the CsPbBr$_3$ sample exhibit a monotonically decreasing trend with increasing kinetic energy, which depends on the sample thickness; in particular, for a given initial kinetic energy of the primary electrons, the reflection is higher for thicker samples and is almost zero at 50 keV, i.e. all electrons are transmitted. At 5 keV, the reflection rate for the 200 nm sample is more than twice as high as for the 50 nm sample ($\eta_B$ = 0.52 compared to $\eta_B$ = 0.22, respectively). The intensities of the elastic peaks in Fig. \ref{EELS} show the decisive contribution of the elastic collisions to the backscattering coefficient.

\subsection{Secondary electrons}

Electron beams that hit a material can also trigger the emission of secondary electrons by depositing their kinetic energy in the target. These electrons are emitted by the atoms of the solid as a result of inelastic interactions with the target, which may involve either the primary beam or other secondary electrons. The yield of transmitted or reflected secondary electrons recorded by the detector is influenced by all kinds of inelastic interactions with individual and collective charges, such as Auger decay, phonons, plasmons, trapping and charge density waves (CDW). Secondary electron emission is the primary contrast mechanism in an SEM and in SE imaging within the TEM. Secondary electron generation can also contribute to the charging of a sample, which has implications on electron beam-induced damage mechanisms \cite{Atomic_Resolution_Secondary,sec_induced_mech,Se_Damage_cambridge_2022,Egerton_2021}.
\begin{figure}[hbt!]
\includegraphics[width=0.8\linewidth]{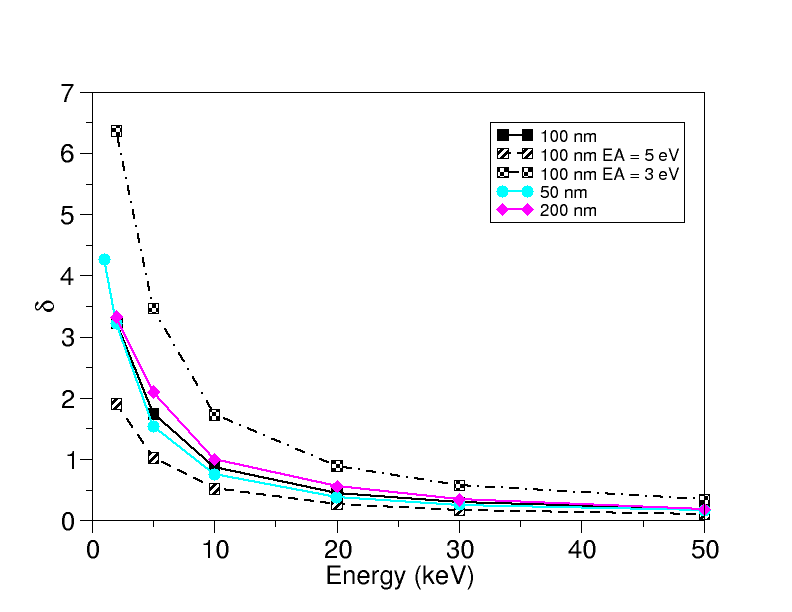}
\caption{Reflection yield of the secondary electrons ($\delta$) as a function of the kinetic energy of the primary beam (keV).
The results are shown for three different sample thicknesses and electron affinities: 50 nm (cyan-coloured solid line), 100 nm (black solid line) and 200 nm (magenta-coloured solid line) for EA = 4.1 eV (pristine material). For the 100 nm layer, results with EA = 3.0 eV (black dash-dotted line) and 5.0 eV (black dashed line) are also included.
}
\label{Sec_reflection}
\end{figure}
In this context, the SE yield is defined as the ratio between the number of secondary electrons (reflected or transmitted with a kinetic energy below 50 eV, although secondary electrons above this nominal limit can also be found) and the total number of primary electrons. The secondary yield is thus defined per incident particle.\\
\indent In Fig. \ref{Sec_reflection} we show the yield of reflected secondary electrons $\delta$ determined for different initial kinetic energies of the primary electron beam of $1, 2, 5, 10, 20, 30, 50$ keV for thicknesses of 50 (cyan line), 100 (black lines) and 200 nm (magenta line) for EA = 4.1 eV (pristine material). To investigate the dependence of the yield on the EA, which may depend on surface termination and material history, calculations were also performed for EA = 3 eV (black dash-dotted line) and 5 eV (black dashed line) and 100 nm thickness.\\
\indent The reflection yield of secondary electrons decreases with increasing initial kinetic energy, from which we can conclude that low-energy primary electrons have a greater chance of producing the escaping secondary electrons immediately below the target surface. The secondary reflection at low kinetic energy of the incident electrons shows a behaviour that is almost independent of the sample thickness, in contrast to the backscattering coefficient of the reflected primary electrons. In addition, the EA plays a fundamental role in increasing the production of secondary electrons. At low kinetic energy of the primary beam, an EA = 3 eV leads to increased production of secondary electrons, while an EA = 5 eV attenuates them. This is particularly evident at the lowest energies (e.g. 2 and 5 keV) and confirms that impinging electrons with lower kinetic energy have a higher probability of emitting secondary electrons. At odds, the electrons penetrate deep into the sample at the high energy of the primary beam (50 keV) and the generated secondary electrons cannot escape at the vacuum-target interface due to the numerous inelastic scatterings they suffer on their way out of the solid; in this case, the different EAs have little influence on the secondary reflection yield.
\begin{figure}[h!]
    \centering
\includegraphics[width=0.32\textwidth]{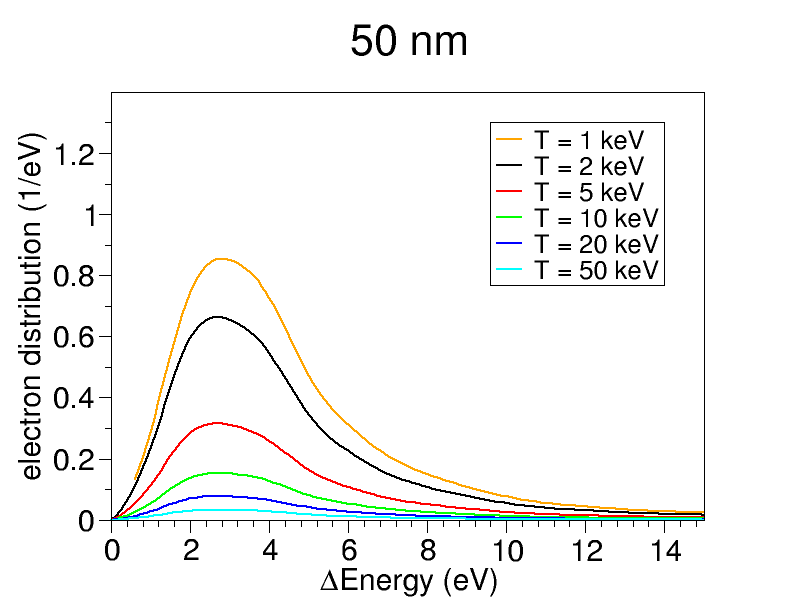}
\includegraphics[width=0.32\textwidth]{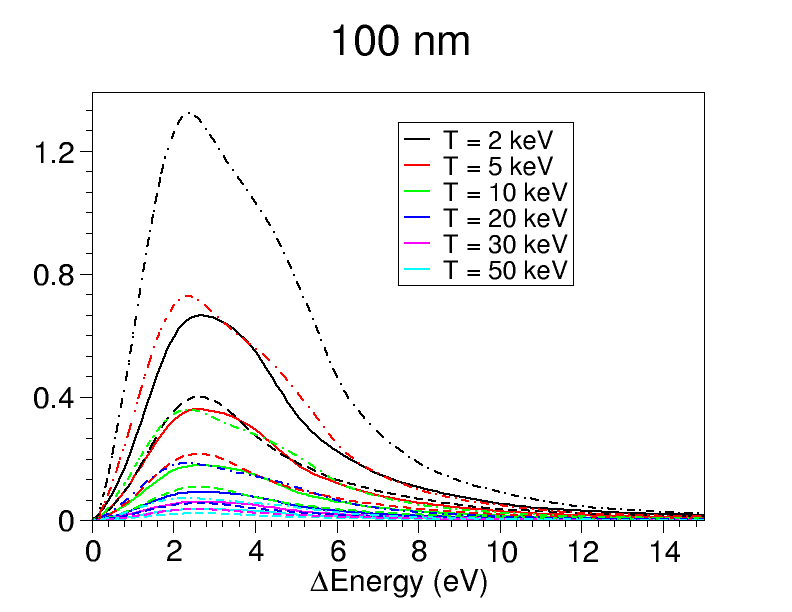}
\includegraphics[width=0.32\textwidth]{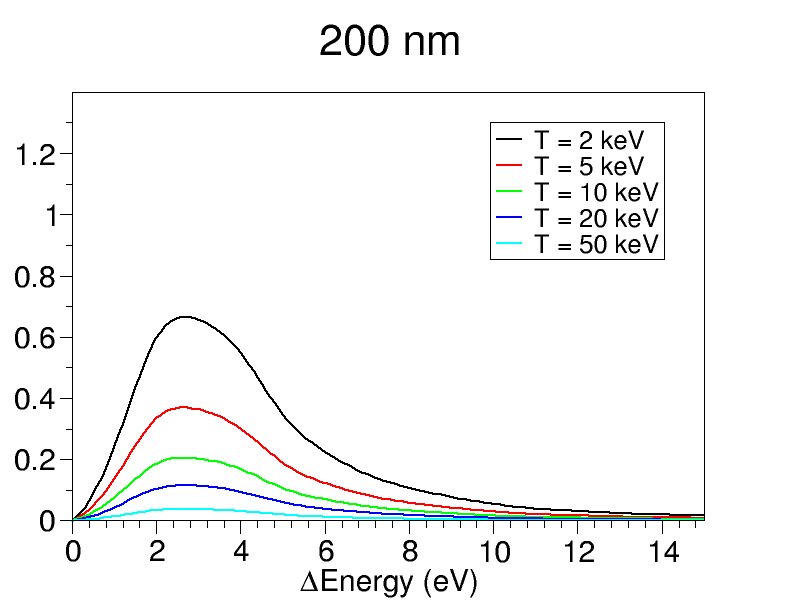}
\caption{Spectral distribution (with respect to the vacuum energy $\Delta \mathrm{Energy}= E-E_\mathrm{vac}$, eV) of the reflected secondary electrons, parameterised for different initial kinetic energies of the primary electrons for three-layer thicknesses of 50 (left), 100 (centre) and 200 nm (right). For the 100 nm thick layer, the electron distribution was also calculated for an EA of 3.0 eV (dotted--dashed lines) and 5.0 eV (dashed lines), while the solid lines represent the pristine CsPbBr{$_3$} (EA = 4.1 eV). The coloured lines represent the different initial kinetic energies of the primary beam of 1 (orange lines), 2 (black lines), 5 (red lines), 10 (green lines), 20 (blue lines), 30 (magenta lines) and 50 keV (cyan lines).}
\label{Electron_distribution_reflection}
\end{figure}

In Fig. \ref{Electron_distribution_reflection} the secondary electron energy ($ \Delta \mathrm{Energy} = E- E_\mathrm{vac}$, with $E_\mathrm{vac}$ the near-surface vacuum energy) spectra are shown as a function of the initial kinetic energy of the primary beam equal to 1 (orange solid line), 2 (black solid line), 5 (red solid line), 10 (green solid line), 20 (blue solid line) and 50 keV (cyan solid line) for three different thicknesses of 50 (left), 100 (centre), 200 nm (right). The resulting secondary electrons are distributed over a range of 25 eV with a sustained peak of around 2.6 eV for all layer thicknesses and kinetic energies of the primary beam. These lineshapes are consistent with the secondary kinetic energy spectra obtained by real-time TDDFT (RT-TDDFT) calculations for graphene, where a peak position at 3.5 eV \cite{RT_TDDFT_doi:10.1021/acs.nanolett.4c00356} with an EA=4.56 eV was found, for semiconductor materials such as TiN, VN, GaAs, InAs, InSb, PbS using the reverse MC approach \cite{Ding_sec}, and with the analytical expression for metals originally developed by Chung--Everhart \cite{chung_1974}, which relates the position of the secondary electron peak to the EA. The Chung--Everhart expression can also be used to estimate grossly the peak distributions for non-metals.
\begin{figure}[hbt!]
\includegraphics[width=0.8\linewidth]{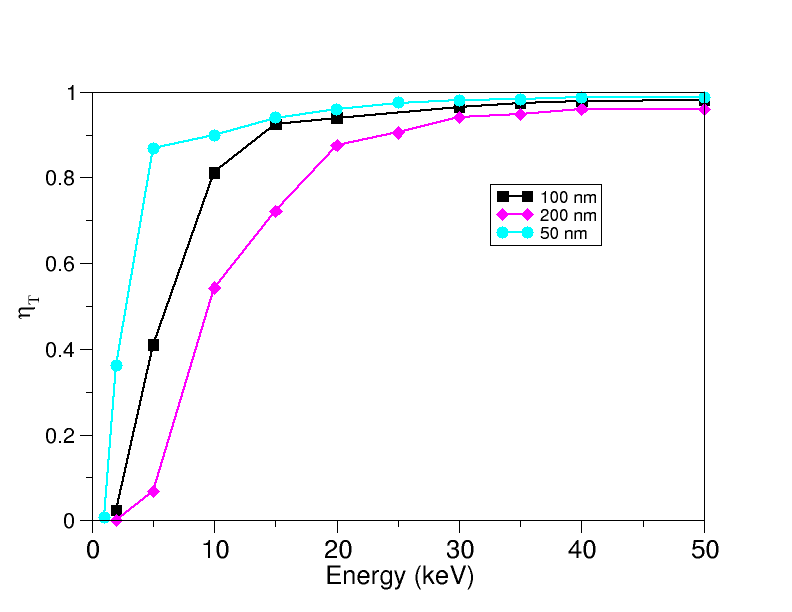}
\caption{\label{graph} Transmission coefficient ($\eta_T$) for the primary electrons as a function of the initial electron kinetic energy (keV).
The results are shown for three different sample thicknesses:  50 nm (cyan solid line), 100 nm (black solid line) and 200 nm (magenta solid line).
}
\label{ePrimary_transm}
\end{figure}
Considering the CsPbBr$_3$ EA = $4.1$ eV, the Chung--Everhart formula provides a peak position measured from the vacuum level at 1/3 of the EA, i.e. $4.1/3\simeq 1.36$ eV ($\simeq$ 1.3 eV below our estimate). Nevertheless, in our simulations, we confirm a red (blue) shift of the peak position by 0.3 eV for an EA of 3.0 eV (5.0 eV). At a thickness of 100 nm (centre panel of Fig. \ref{Electron_distribution_reflection}), the positively charged sample (EA = 3 eV, dotted-dashed line) enhances the intensity of the spectra of the kinetic energy of the secondary electrons, while the EA=5 eV (corresponding to negatively charged samples, dashed line) attenuates it.\\
\indent 
The conclusions that can be drawn from these results for the reflected secondary electron yield are as follows: (i) The more intense secondary electron distributions are found for low kinetic energies (5-10 keV) of the primary electrons rather than for higher energies (e.g. 50 keV). (ii) The EAs influence the distributions of the secondary electron yield mainly at low kinetic energies of the primary electrons (with a maximum at $2.6$ keV). (iii) The distributions are not significantly influenced by the thickness of the layer.



\subsection {Transmission of the primary and secondary electrons.}\label{Transmission_section}

The analysis carried out for the backscattered electrons is repeated for the primary and secondary electron transmission. In Fig. \ref{ePrimary_transm}, the transmission coefficient ($\eta_T$) is plotted against the initial kinetic energies of the primary beam for three thicknesses. The number of transmitted primary electrons per incident particle is always less than 1 at low kinetic energy, as a large fraction of the incident electrons are reflected. However, the primary transmission coefficient saturates at a higher kinetic energy ($> 30$ keV) for all investigated samples. We also note that the transmission coefficient decreases with increasing thickness at all energies. In particular, for the 100 nm thick layer (black solid line), the transmission coefficient of the primary electrons shows a behaviour between 50 (cyan solid line) and 200 nm (magenta solid line). In the case of 50 nm (cyan solid line), the primary electrons are almost completely transmitted at an incident energy of 5 keV. We emphasise that the spectra and yield of the secondary electrons, which are typically orders of magnitude greater in number than the primary electrons, have a better signal-to-noise ratio than the corresponding magnitudes of the primary electrons.

\begin{figure}[hbt!]
\includegraphics[width=0.8\linewidth]{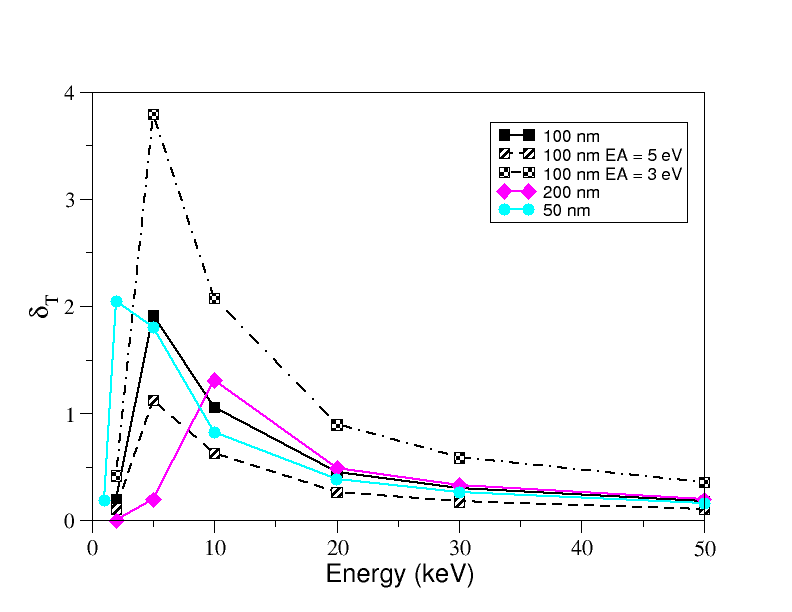}
\caption{\label{graph} Secondary electron transmission yield ($\delta_\mathrm{T}$) as a function of the initial kinetic energy of the primary beam (keV).
The results are shown for three different sample thicknesses and electron affinities: pristine material (EA= 4.1 eV), 50 nm (cyan solid line), 100 nm (black solid line) and 200 nm (magenta solid line). The results for a depth of 100 nm with EA = 3.0 eV (black dashed line) and 5.0 eV (black dashed line) are also shown.
}
\label{eSec_Trans}
\end{figure} 
The transmission coefficient of the secondary electrons as a function of the primary beam kinetic energy reported in Fig. \ref{eSec_Trans} shows a maximum, whose position and intensity depend on the thickness: 2 keV for 50 nm (cyan solid line), 5 keV for 100 nm (black solid line) and 10 keV for 200 nm (magenta solid line). 
\begin{figure}[hbt!]
\centering
\includegraphics[width=0.32\textwidth]{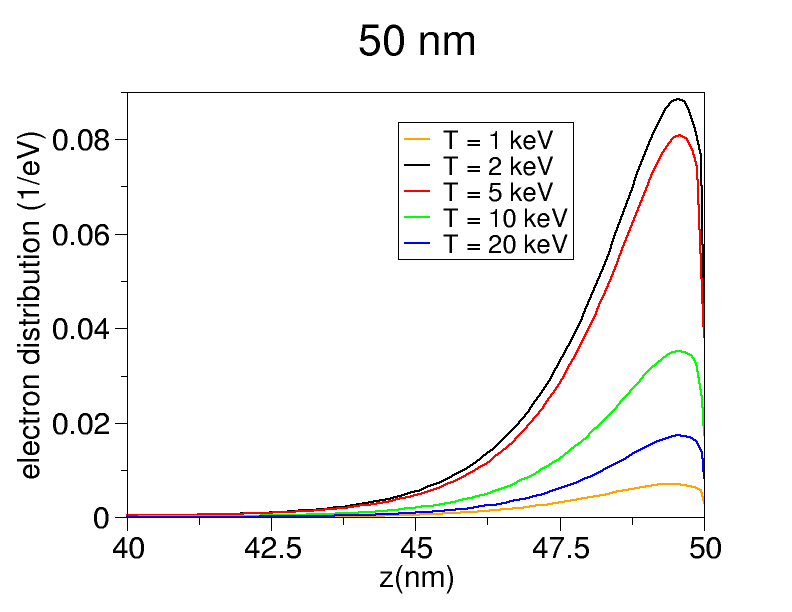}
\includegraphics[width=0.32\textwidth]{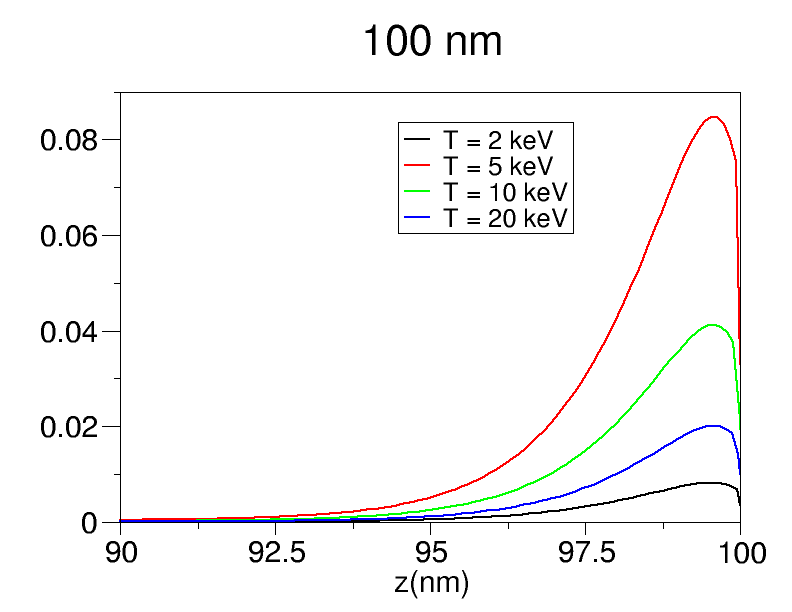}
\includegraphics[width=0.32\textwidth]{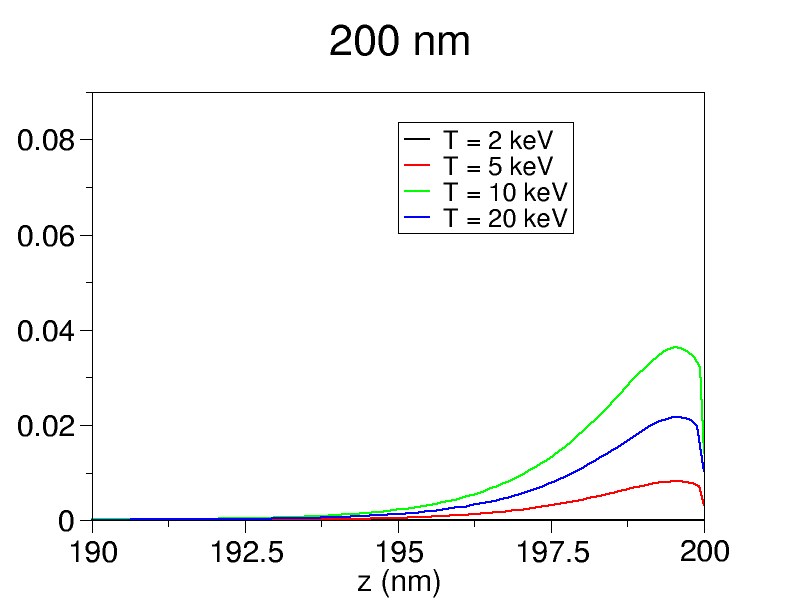}
\caption{\label{trans_sec_depth} Distribution of transmitted secondary electrons as a function of the origin depth (along $z$-axis, in nm), parameterised for different initial primary electron energies for slab thicknesses of 50 (left), 100 (centre) and 200 (right) nm, respectively. The initial kinetic energies of the primary electron beam are T=1 keV (orange lines), 2 keV (black lines), 5 keV (red lines), 10 keV (green lines) and 20 keV (blue lines).}
\end{figure}
We also find that the intensity of the maximum of $\delta_\mathrm{T}$ decreases with increasing thickness and the EA also significantly affects the intensity of the transmitted secondary electrons; in particular the higher the EA, the smaller $\delta_\mathrm{T}$.
We attribute this result to the fact that the only secondary electrons that can escape are those that originate near the vacuum interface (opposite the entrance surface).
This behaviour can be interpreted in the light of Fig. \ref{trans_sec_depth}, where we plot the distribution of transmitted secondary electrons as a function of origin depth. Fig. \ref{trans_sec_depth} shows that the number of generated secondary electrons initially increases as the energy of the primary electrons increases. However, this trend reverses as soon as the energy exceeds a certain threshold value, which is determined by the thickness of the film. This leads to a saturation point at which a considerable number of electrons from the primary beam pass through the film without much interaction and no longer contribute to the production of secondary electrons.  This is also reflected in the thickness-dependent plateau in Fig. \ref{ePrimary_transm} for the transmission of primary electrons ($>$ 30 keV). Moreover, we find that the maximum secondary electron transmission occurs at a depth of about $0.75$ nm  along the direction of incidence from the interface to the vacuum. Finally, we also note that the generation of secondary electrons is filtered out at a low initial kinetic energy of the primary beam, e.g. the 2 keV distribution is most intense in a 50 nm thick layer, while it is attenuated in the 100 nm thick layer and suppressed in the 200 nm thick layer. 

\begin{figure}[hbt!]
 \includegraphics[width=0.32\textwidth]{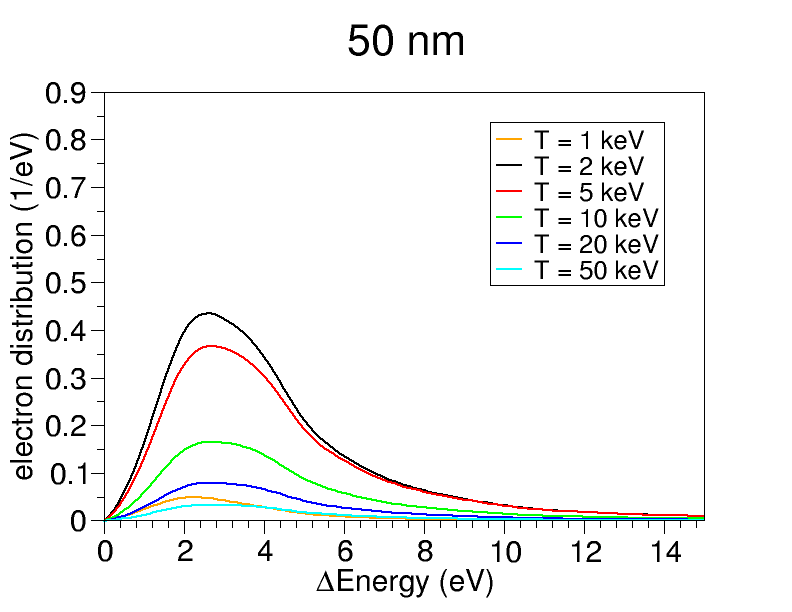}
\includegraphics[width=0.32\textwidth]{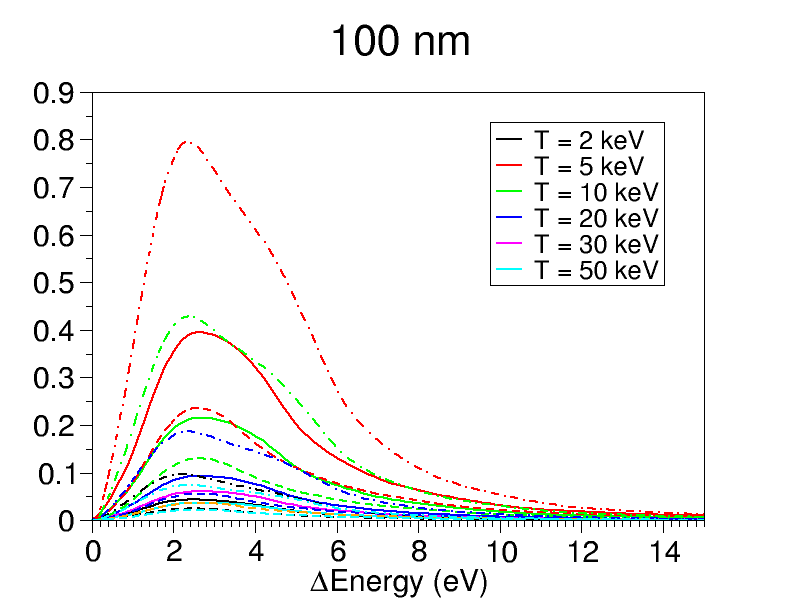}
 \includegraphics[width=0.32\textwidth]{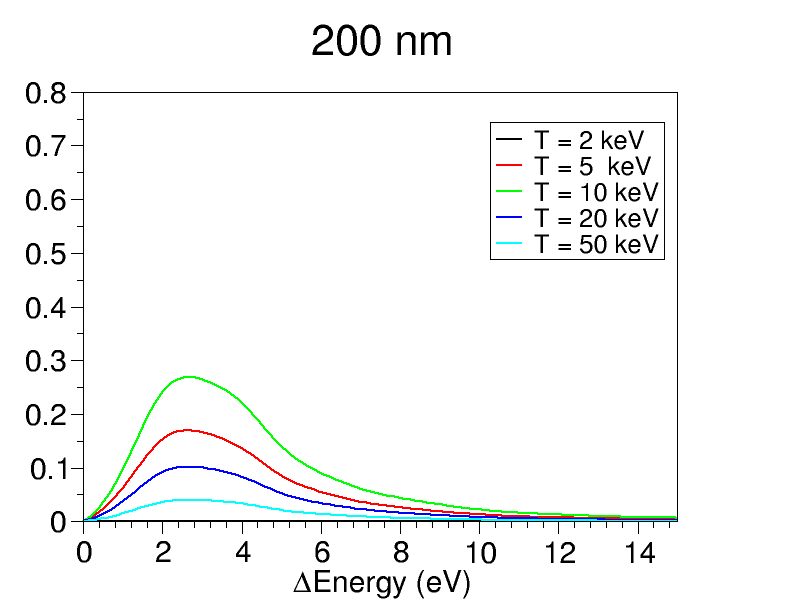}
\caption{\label{transport_sec_distribution} Energy spectra of the transmitted secondary electrons as a function of the kinetic energy of emission (with respect to the vacuum energy $\Delta \mathrm{Energy}= E-E_\mathrm{vac}$, eV) for different initial primary electron energies for the three slab thicknesses, namely 50 (left), 100 (centre) and 200 (right) nm. For the 100 nm slab, the electron distribution was also calculated with an EA of 3.0 eV (dotted-dashed lines) and 5.0 eV (dashed lines), while for the original CsPbBr{$_3$} the EA is 4.1 eV (solid lines). The initial kinetic energies of the primary beam are $T=$ 1 keV (orange lines), 2 keV (black lines), 5 keV (red lines), 10 keV (green lines), 20 keV (blue lines), 30 keV (magenta lines) and 50 keV (cyan lines).}
\end{figure}
Finally, Fig. \ref{transport_sec_distribution} shows the energy distributions of the transmitted secondary electrons for different parametrised layer thicknesses. Similar to the behaviour of the reflected secondary electrons, the energy spectra of the transmitted secondary electrons also exhibit a persistent peak around 2.6 eV, independent of the thickness and kinetic energy of the primary beam, with a tail reaching up to 20 eV. EA of 3 (5) eV increases (attenuates) the intensity of the peaks and produces a small red (blue) shift in the peak position, as shown in the centre panel of Fig. \ref{transport_sec_distribution}. The EA affects the transmission yield of secondary electrons, even at a kinetic energy of 50 keV. 

\section{Discussion and conclusions}

In this study, we have theoretically investigated the scattering processes of high-energy primary and secondary electrons expected from scanning transmission electron microscopy measurements of the perovskite CsPbBr$_{3}$. To this end, we used MC simulations in combination with first-principles LR-TDDFT calculations of the electron loss function to determine the inelastic mean free path of bulk CsPbBr$_3$. \\
\indent We have focused in particular on analysing the reflection and transmission properties of backscattered and secondary electrons by calculating the spectra and yields of the emitted secondary electrons in thin films of CsPbBr$_3$.\\
\indent Our results show that the backscattering coefficient of the primary electrons decreases monotonically with kinetic energy and increases with film thickness, while the transmission coefficient is larger for thinner films and approaches unity for initial kinetic energies of the primary beam above 30 keV for all thicknesses investigated.
In the reflection and transmission of secondary electrons, the EA plays a critical role in producing these outgoing electrons, although its influence diminishes at higher energies.\\
\indent
In terms of transmission, the observed peaks depend on both the thickness of the material and the initial kinetic energy of the primary electrons. Outgoing secondary electrons are most likely to be emitted when they are generated within the last 10 nm of the nanolayer. Therefore, the kinetic energy of the incoming electrons must be carefully balanced; it should be high enough to generate secondary electrons without releasing too much energy and low enough to prevent the generation of secondary electrons too deeply within the sample.\\
\indent
These results contribute to a better understanding of the transmission and reflection characteristics of primary and secondary electrons, ultimately aiding in the quantitative interpretation of intensity in STEM experimental measurements.

\begin{acknowledgments} 
This work was supported by the Ada Lovelace Centre.
The authors gratefully acknowledge the use of ARCHER2 (via the UK Car-Parrinello Consortium, EP/X035891/1) and the STFC-SCARF HPC facilities. S.T. acknowledges the European Union for funding under grant agreement n° 10104665.
P.E.T. thanks Dr J. Jackson for assistance with the Questaal calculations and Dr I. Scivetti for his help with Archer2.

\end{acknowledgments}

\appendix
\section{Drude-Lorentz method}\label{sec:appendix}

The inelastic scattering cross section is derived from the ELF in the entire spectrum of excitation energies ($\omega$) and momentum transfers $\bf{q}$. While first-principles methods can effectively determine the ELF for energies below 100 eV, performing ab-initio calculations of the ELF over such a broad energy range (up to 400 keV) is often prohibitively expensive due to the enormous number of excited states that must be included in the ELF calculation. To obtain the IMFP in the investigated energy range, we need to rely on experimental NIST X-ray atomic data (TDDFT-NIST) \cite{NIST}.
\begin{figure}[hbt!]
\centering
\includegraphics[width=0.7\linewidth]{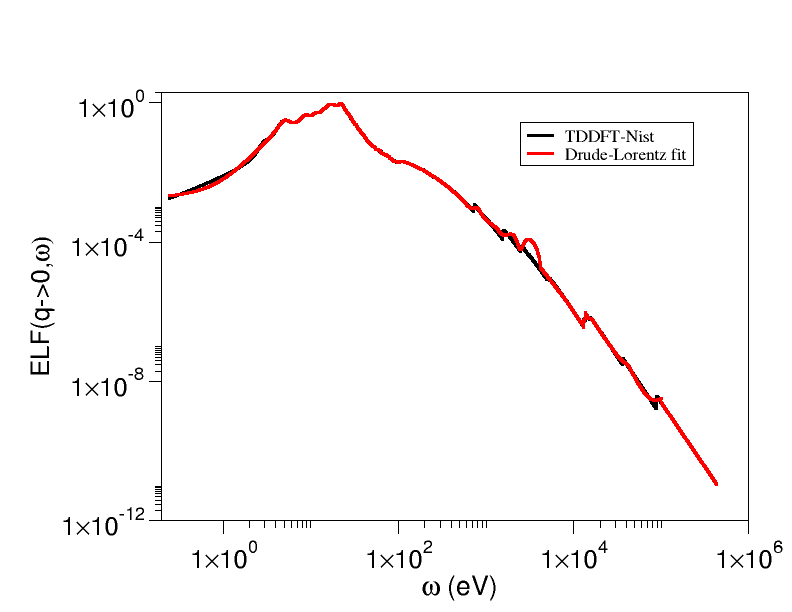}
\caption{Drude-Lorentz fit (black line) of the TDDFT-NIST \cite{NIST} ELF (red line).
}
\label{Drude}
\end{figure}

The TDDFT-NIST ELF was first fitted using the ALC$\_$SUTOR suite \cite{ALC_sutor}, which uses nonlinear least squares minimisation $lmfit$ \cite{lmfit}. We used 34 Drude-Lorentz oscillators with a momentum-dependent broadening parameter $\gamma_i$ as follows:
\begin{equation}
    \mathrm{Im}\left[{\frac{-1}{\overline{\epsilon}(\mathbf{q=0},\omega)}}\right]=\sum_i \frac{A_i\gamma_i\omega}{(\omega^2_i-\omega^2)^2+(\gamma_i\omega)^2},
    \label{eq:fit}
\end{equation}
where $A_i$ represents the intensity of the transition.
The friction damping parameter $\gamma_i$ takes into account the momentum dispersion of the $i$th electronic interband and intraband excitation in the valence and conduction bands, which is characterised by a transition energy $\omega_i$ according to the following formula \cite{PEDRIELLI}:

\begin{equation}
    \begin{split}
        \omega_i(q) &=\sqrt{\omega_i^2+(12/5)E_f\cdot q^2/2+q^4/4}\\
        \gamma_i(q)&=\sqrt{\gamma_i+q^2/2+q^4/4}.
    \end{split}
\end{equation}

To include the electronic excitations of the inner shells in the ELF calculation, we have additionally used 10 Fano functions and 4 power-law functions that accurately reproduce the sharp edges associated with the core excitations.

The result of this fitting procedure is shown in Fig. \ref{Drude}.
For details on different methods to extend the ELF beyond the optical limit, see Ref. \cite{TAIOLI2024100646}.

\bibliography{bibliography}

\begin{thebibliography}{55}
\bibitem [{ese()}]{esempio}%
  \BibitemOpen
  \href@noop {} {}\bibinfo {note} {The plane-wave basis energy cutoff was 520 eV. The Monkhorst-Pack scheme k-points grid sampling was set at $3\times 2 \times 2$ .}\BibitemShut {Stop}%
\end{thebibliography}

\newpage

\end{document}